\documentclass[floats,aps,showpacs,twocolumn,groupedaddress]{revtex4}

\usepackage[dvips]{graphicx}
\usepackage{amsfonts}
\usepackage{amsmath,amssymb}
\usepackage{dsfont}
\usepackage{placeins}
\usepackage{afterpage}
\usepackage{flafter}

\newcommand{\diff}{\text{d}}

\newcommand{\psireg}{\psi_{\text{reg}}}
\newcommand{\psiregmn}{\psi_{\text{reg}}^{mn}}
\newcommand{\psiregmnW}{\psi_{\text{reg},W}^{mn}}
\newcommand{\psich}{\psi_{\text{ch}}}
\newcommand{\Hreg}{H_{\text{reg}}}

\begin{document}

\title{Dynamical tunneling in mushroom billiards}
\author{A.~B\"acker}
\author{R.~Ketzmerick}
\author{S.~L\"ock}
\affiliation{Institut f\"ur Theoretische Physik, Technische
              Universit\"at Dresden, D-01062 Dresden, Germany}
\author{M.~Robnik}
\author{G.~Vidmar}
\affiliation{Center for Applied Mathematics and Theoretical Physics,
              University of Maribor, SI-2000 Maribor, Slovenia}
\author{R.~H\"ohmann}
\author{U.~Kuhl}
\author{H.-J.~St\"ockmann}
\affiliation{Fachbereich Physik, Philipps-Universit\"at Marburg,
              D-35032 Marburg, Germany}

\date{\today}

\begin{abstract}
We study the fundamental question of dynamical tunneling in
generic two-dimensional Hamiltonian systems by considering regular-to-chaotic
tunneling rates.
Experimentally, we use microwave spectra to investigate a mushroom billiard with
adjustable foot height.
Numerically, we obtain tunneling rates from high precision eigenvalues
using the improved method of particular solutions.
Analytically, a prediction is given by extending an approach using a fictitious
integrable system to billiards.
In contrast to previous approaches for billiards, we find agreement with 
experimental and numerical data without any free parameter.
\end{abstract}

\maketitle

Typical Hamiltonian systems have a mixed phase space
in which regular and chaotic motion coexist. While
classically these regions are separated, quantum mechanically
they are coupled by tunneling. This process
has been called ``dynamical tunneling''~\cite{DavHel1981} as it occurs
across a dynamically generated barrier in phase space.
Tunneling has been studied between symmetry related regular
regions (chaos-assisted tunneling)~\cite{BohTomUll1993,FriDor1998,Dem2000,
SteOskRai2001,HenHafBroHecHelMcKMilPhiRolRubUpc2001,BroSchUllEltComb}
and from a single regular region to the chaotic
sea~\cite{HanOttAnt1984,ShuIke1995,SheFisGuaReb2006,FeiBaeKetRotHucBur2006,BaeKetLoeSch2008}.
In contrast to the well understood 1D tunneling through a barrier,
the quantitative prediction of dynamical tunneling
is a major challenge.
Results have been found for specific systems or system
classes only,
e.g.\ recently for 2D quantum maps with an approach
using a fictitious integrable system~\cite{BaeKetLoeSch2008}.
However, a precise knowledge of tunneling rates is of great importance.
Recent examples are spectral statistics in systems with a mixed phase
space \cite{VidStoRobKuhHohGro2007},
eigenstates affected by flooding of regular islands \cite{BaeKetMon200507},
and emission properties of optical micro-cavities \cite{WieHen2008}.

Billiards are an important class of Hamiltonian systems. Classically,
a point particle moves along straight lines inside a domain
with elastic reflections at its boundary.
Quantum-mechanical approaches for dynamical tunneling rates have so far
escaped a full quantitative prediction as they required fitting by a factor
of 6 for the annular billiard \cite{FriDor1998} and by a factor of 100 (see below) for the
mushroom billiard \cite{BarBet2007}.

 \begin{figure}[b]
  \begin{center}
    \includegraphics[angle = 0, width = 85mm]{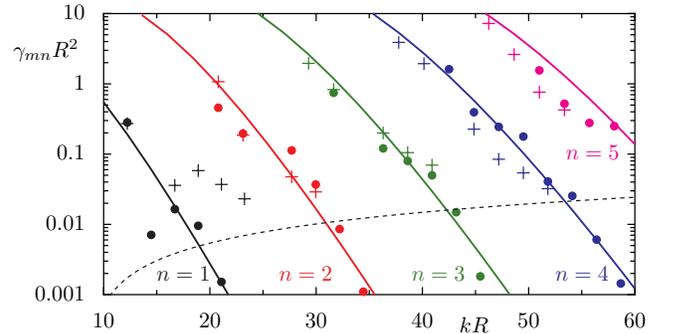}
     \caption{(color online) Dynamical tunneling rates from the regular region
           to the chaotic sea for quantum numbers $n\leq 5$ vs $kR$
           for a mushroom billiard ($a/R=10/19$):
           theoretical predictions (connected by solid lines)
           from Eq.~\eqref{eq:final_formula}, experimental results
           (crosses), and numerical data (dots).
           The dashed line denotes the lower
           limit of the experimental resolution.
           }
     \label{fig:rates_exp}
   \end{center}
 \end{figure}

\begin{figure}[tbp]
   \begin{center}
    \includegraphics[angle = 0, width = 85mm]{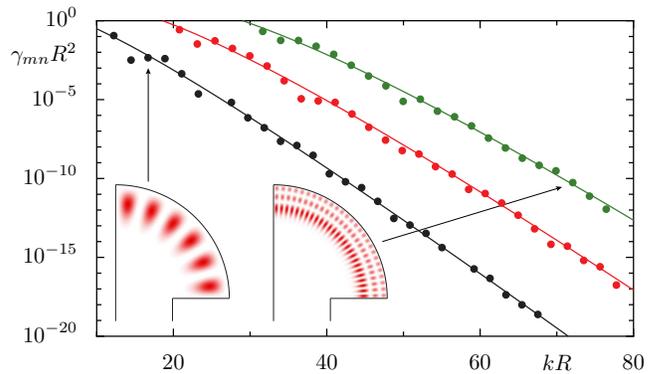}
     \caption{(color online) Tunneling rates from regular states
           with quantum numbers $n=1,$ $2,$ and $3$ vs $kR$ for $a/R=0.5$
           comparing prediction Eq.~\eqref{eq:final_formula}
           (connected by solid lines) and numerical data (dots).
           The insets show the regular eigenfunctions $\psireg^{12,1}(x,y)$
           and $\psireg^{54,3}(x,y)$.
          }
     \label{fig:rates_low_E}
   \end{center}
 \end{figure}

In this paper we present a combined experimental, theoretical,
and numerical investigation of dynamical tunneling rates
in mushroom billiards \cite{Bun2001}, which are of great current
interest \cite{AltMotKan2005,TanShu2006,DieFriMisRicSch2007,VidStoRobKuhHohGro2007,BarBet2007}
due to their sharply divided phase space.
Experiments are performed using a microwave cavity.
Extending the approach using a fictitious integrable system
\cite{BaeKetLoeSch2008} to billiards, we find quantitative agreement in
the experimentally accessible regime, see Fig.~\ref{fig:rates_exp},
without a free parameter. In addition,
numerical computations verify the predictions
over 18 orders of magnitude
with errors typically smaller than a factor of 2,
see Fig.~\ref{fig:rates_low_E}.
The theoretical approach thus provides unprecedented agreement
for tunneling rates in billiards.

\begin{figure}[b]
   \begin{center}
    \includegraphics{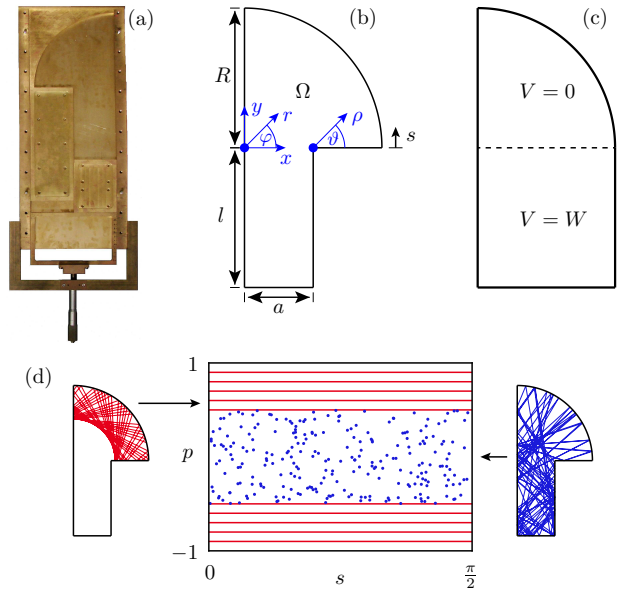}
     \caption{(a) Experimental desymmetrized mushroom billiard
              with radius $R= 19$\,cm, foot width $a= 10$\,cm and foot height
              $l=0\dots 25.7$\,cm.
              The antenna is located 4\,cm below the top and has a distance
              of 0.75\,cm from the vertical wall.
              (b) Schematic picture showing the coordinate systems
              used in the theoretical derivation.
              (c) Auxiliary billiard $\Hreg^W$.
              (d) Phase-space portrait at the quarter circle boundary
              (relative tangential momentum $p$ vs arclength $s$)
              showing regular and chaotic regions with
              illustrations of trajectories.
              }
     \label{fig:mushroom}
   \end{center}
 \end{figure}

We consider the desymmetrized mushroom billiard, i.e. the 2D
autonomous system
$H(\mathbf{p},\mathbf{q})=\mathbf{p}^2/2M+V(\mathbf{q})$ shown in
Fig.~\ref{fig:mushroom}b, characterized by the radius of the quarter
circle $R$, the foot width $a$ and the foot height $l$. The
potential is zero inside the domain $\Omega$ and infinite outside.
Classically one has regular and chaotic dynamics as visualized by
the phase-space portrait in Fig.~\ref{fig:mushroom}d.
Quantum mechanically the billiard
is described by the time-independent Schr\"odinger equation
$-\Delta\psi(\mathbf{q})=E\psi(\mathbf{q})$ with Dirichlet boundary
conditions at $\partial\Omega$, using the natural units $2M=\hbar=1$.
The eigenstates can be classified as being either mainly
regular or mainly chaotic, depending on the phase space region
on which they concentrate.

The initial tunneling decay of a purely (unperturbed) regular state,
is, according to Fermi's golden rule, described by a rate
\begin{equation}
\label{eq:FGR}
 \gamma = 2\pi \langle |v|^2 \rangle \rho_{\text{ch}},
\end{equation}
where $\langle |v|^2 \rangle$ is the averaged squared matrix element
between this regular state and the chaotic states of similar energy.
According to Weyl's formula the density of chaotic states is
$\rho_{\text{ch}} \approx A_{\text{ch}}/4\pi$, where
$A_{\text{ch}}=la+[R^2\arcsin{(a/R)}+a\sqrt{R^2-a^2}]/2$ is
the area of the billiard times the fraction of the chaotic
phase-space volume.

Under variation of the foot height~$l$
(and for not too large $\rho_{\text{ch}}$ \cite{BaeKetMon200507})
avoided crossings of this regular state with different
chaotic states are observable. The splittings
$\Delta E = 2|v|$ determine individual matrix elements
for different billiards with varying $\rho_{\text{ch}}$.
We therefore extend the average in Eq.~\eqref{eq:FGR} also over
$\rho_{\text{ch}}$ relying on the assumption
that the regular-to-chaotic tunneling rate $\gamma$
is a local property of the hat region,
where the regular phase-space component resides,
and thus does not depend on the foot height. This leads to
\begin{eqnarray}
\label{eq:ensembleaverage}
  \gamma = \langle |\Delta E|^2 A_{\text{ch}}/8\rangle ,	
\end{eqnarray}
where $A_{\text{ch}}$ in the experiment varies
from 0.5 to 1.2 when increasing $l$ from
0 to 25.7\,cm.

\begin{figure}[tb]
   \begin{center}
    \includegraphics[width=\columnwidth]{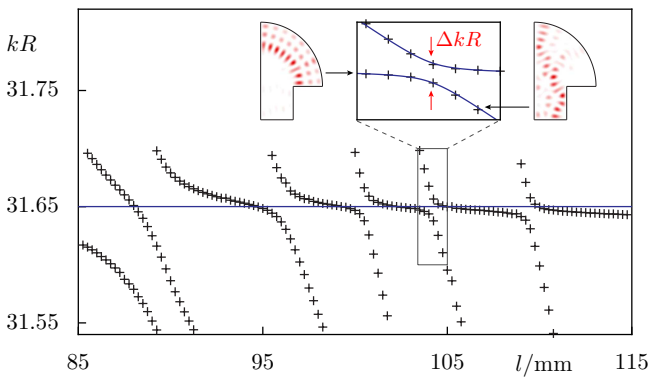}
     \caption{(color online)
     Part of the evaluated experimental resonance spectra of the mushroom
     microwave billiard vs foot height $l$.
     The horizontal line corresponds to the eigenfrequency of $\psireg^{18,3}$,
     the crosses mark the extracted resonances.
     The inset shows a magnification of one avoided crossing, including
     a hyperbola fit (solid line),
     and the numerically obtained regular and chaotic states involved in
     the crossing. 
     }
     \label{fig:spaghetti}
   \end{center}
 \end{figure}

Fig.~\ref{fig:mushroom}a shows the mushroom billiard
used in the microwave experiment. Spectra have been taken as a function of the
foot height $l$ of the mushroom in the frequency regime 3.0 to 13.5 GHz,
corresponding to values of $kR$ between 11.9 and 53.8.
Fig.~\ref{fig:spaghetti} shows part of the obtained spectra in a small $kR$
window. As the energy of the regular states of the quarter
circle do not depend on the foot height, they appear
as straight horizontal lines, whereas the chaotic states are shifted
to lower energies with increasing foot height $l$, reflecting the
increasing density $\rho_{\text{ch}}$ of chaotic states.

For each of the regular states, see Eq.~\eqref{eq:regstates},
with radial quantum numbers $n$ between $1$ and $5$, and azimuthal
quantum numbers $m$ even between $8$ and $32$ all accessible splittings at
avoided crossing $\Delta k$ have been determined by means of a
hyperbola fit (Fig.~\ref{fig:spaghetti}). From this we get
the energy splittings $\Delta E = 2k \Delta k$
of the corresponding quantum system and by averaging
over all avoided crossings, Eq.~\eqref{eq:ensembleaverage},
deduce the tunneling rates $\gamma_{mn}$
from regular states $(m,n)$ to the chaotic sea. Apart from the
results for $n=1$ they are in very good agreement with the
theoretical prediction, Eq.~\eqref{eq:final_formula}, derived below.

The experimental resolution of avoided crossings is limited by the
line widths of the resonances caused by wall absorption and antenna
coupling. In the studied frequency regime the line widths were about
$\Delta\nu_w=0.01$\,GHz, corresponding to a $\Delta k_wR \approx 0.004$.
From the hyperbola fit of the avoided
crossings all splittings $\Delta kR$ larger than  $0.1\Delta k_wR$
could still be resolved, corresponding to
tunneling rates $\gamma$ between 0.001 and 0.024, see Fig.~\ref{fig:rates_exp}
(dashed line). 
Another complication
is caused by the antenna giving rise to an additional splitting
\cite{VidStoRobKuhHohGro2007}, which
is proportional to the product of the involved wave
functions $|\psi(\mathbf{q}_a)|$ at the antenna position $\mathbf{q}_a$.
For the rightmost three data points
for $n=1$, see Fig.~\ref{fig:rates_exp},
$|\psireg^{m1}(\mathbf{q}_a)|$ is particularly large, which
is probably the explanation for the deviations between experiment
and theory observed in these cases.

Now we derive a formula for tunneling rates in billiards.
This faces a general problem:
The matrix element $v$, appearing in Eq.~\eqref{eq:FGR},
cannot be calculated from the nominally regular and chaotic eigenstates of $H$,
as they have small, but still too large, admixtures of the other type of
states. 
Instead, we determine $v$ by introducing a fictitious regular 
billiard system $\Hreg$ with purely regular eigenstates,
extending an approach for 1D quantum maps~\cite{BaeKetLoeSch2008}.
$\Hreg$ has to be chosen such that its classical dynamics resembles
the classical motion corresponding to $H$ within the regular region
as closely as possible.
The eigenstates $\psireg$ of $\Hreg$ are localized in the regular
region and continue to decay into the chaotic sea.
With states $\psich$ living in the chaotic
region of phase space, we use in analogy to Ref.~\cite{BaeKetLoeSch2008}
for the coupling matrix element
\begin{equation}
\label{eq:coupmatel}
 v = \int_{\Omega}\psich^*(x,y)(H-\Hreg)\psireg(x,y)
                   \,\diff x \,\diff y.
\end{equation}
Eqs.~\eqref{eq:FGR} and \eqref{eq:coupmatel}
define our approach for determining dynamical tunneling rates in billiards.
Note, that it requires the determination of reasonably good $\Hreg$ and
$\psireg$, which for a general billiard is a difficult task.

We will now apply this approach to the desymmetrized mushroom billiard,
see Fig.~\ref{fig:mushroom}b.
We set $R=1$ in the following analysis.
A natural choice for the
regular system $\Hreg$ is the quarter-circle billiard with its eigenstates
\begin{eqnarray}
\label{eq:regstates}
 \psiregmn(r,\varphi) = N_{mn} J_m\left(j_{mn} r\right)\sin(m\varphi),
\end{eqnarray}
in polar coordinates $(r,\varphi)$.
They are characterized by the radial ($n=1,2,\dots$)
and the azimuthal ($m=2,4,\dots$) quantum numbers.
Here $J_m$ denotes the $m$th Bessel function, $j_{mn}$ the $n$th root of
$J_m$, $N_{mn}=\sqrt{8/\pi}/J_{m-1}(j_{mn})$ the normalization,
and $E_{mn}=j_{mn}^{2}$ is the eigenenergy.

Evaluating Eq.~\eqref{eq:coupmatel} leads for $y\leq 0$ to the
undefined product of $H-\Hreg=-\infty$ and $\psiregmn=0$.
We therefore introduce the auxiliary billiard $\Hreg^W$, see
Fig.~\ref{fig:mushroom}c, with a large but finite potential
$V(x,y\leq0)=W \gg E$.
We evaluate Eq.~\eqref{eq:coupmatel} in the
limit $W\to\infty$, where $\Hreg^W$ approaches $\Hreg$, leading to
\begin{eqnarray}
\label{eq:coupmatel2}
 v & = & \lim_{W\to\infty} \int_{0}^{a}\diff x \int_{-l}^{0}
           \diff y\,\psich(x,y)(-W)\psiregmnW(x,y)\nonumber\\
   & = & -\int_{0}^{a} \diff x\, \psich(x,y=0)\,\partial_y \psiregmn(x,y=0),
\end{eqnarray}
where we performed the $y$-integration on
$\psiregmnW(x,y)=\psiregmnW(x,y=0)\exp(\sqrt{W-E_{mn}} y)$,
following from the Schr\"odinger equation
for $y<0$ and the continuity at $y=0$.
Furthermore we used
$\partial_y \psiregmnW(x,y=0) = \sqrt{W-E_{mn}} \psiregmnW(x,y=0)$,
following from the continuity of the derivative at $y=0$,
and we replaced $\lim_{W\to\infty} \sqrt{W} \exp(\sqrt{W-E_{mn}} y)$
by a Dirac delta function. Below we use $\partial_y \psiregmn(x,0)
= m N_{mn} J_m(j_{mn}x)/x$.

For the chaotic states $\psich(x,y)$ we employ
a random wave description \cite{Ber1977b},
which has recently been extended to
systems with a mixed phase space \cite{BaeSch2002a}.
While this describes the behavior inside the billiard
accurately, it would not include the effect of the boundary,
e.g.\ near the corner, where the main contribution
of the integral in Eq.~\eqref{eq:coupmatel2} arises.
We extend a boundary-adapted random wave model~\cite{Ber2002}
to the case of a corner with angle $3\pi/2$
using basis states with Dirichlet boundary conditions~\cite{Leh1959},
\begin{eqnarray}
\label{eq:chaoticstate}
 \psich(\rho,\vartheta) \approx \sqrt{\frac{8}{3A_{\text{ch}}}}
        \sum_{s=1}^{\infty} c_{s} J_{\frac{2s}{3}}\left(\sqrt{E}\rho\right)
                            \sin\left(\frac{2s}{3}\vartheta\right),
\end{eqnarray}
where the polar coordinates $(\rho,\vartheta)$ at the corner
are related to $(x,y)$ by $x=a+\rho\cos(\vartheta)$
and $y=\rho\sin(\vartheta)$ (see Fig.~\ref{fig:mushroom}b).
The coefficients $c_{s}$ of this ensemble are
independent Gaussian random variables with
$\langle c_{s} \rangle = 0$ and
$\langle c_{s} c_{t}\rangle = \delta_{s,t}$.
The normalization is chosen such that
$\langle|\psich(\rho,\vartheta)|^2\rangle = 1/A_{\text{ch}}$
holds far away from the corner.
Note, that (i)
we do not require these chaotic states to decay into the regular island,
as Eq.~\eqref{eq:coupmatel2} is an integral
along a line of the billiard where the phase space is fully chaotic,
and that (ii)
near the boundary, but away
from the corner, one recovers the behavior $1-J_0(2 k |x|)$
\cite{BaeSchSti1998,Ber2002}.
Inserting  Eq.~\eqref{eq:chaoticstate} for $\vartheta=\pi$
into Eq.~\eqref{eq:coupmatel2}
one can determine the averaged squared matrix element,
$\langle |v|^2 \rangle$,
and with Eq.~\eqref{eq:FGR} one gets
\begin{equation}
\label{eq:formula_int}
 \gamma_{mn} = m^2 N_{mn}^2
 \sideset{}{^\prime}\sum_{s=1}^{\infty} \left[\int\limits_{0}^{a}
 \frac{\diff x}{x} J_m(j_{mn}x) J_{\frac{2s}{3}}(j_{mn}[a \text{-} x])\right]^2,
\end{equation}
where the sum over $s$ excludes all multiples of $3$,
which is indicated by the prime.
The remaining integral can be solved
analytically, leading to
\begin{equation}
\label{eq:final_formula}
 \gamma_{mn} = \frac{8}{\pi}\sideset{}{^\prime}\sum_{s=1}^{\infty}
               \frac{J_{m+\frac{2s}{3}}(j_{mn}a)^2}{J_{m-1}(j_{mn})^2}
\end{equation}
for the tunneling rates from any regular state $\psiregmn$ in the
mushroom billiard. The sum has its
dominant contribution for $s=1$ and using $s\leq 2$ is sufficiently
accurate.

It is worth to remark that a very plausible estimate of the
tunneling rate is given by the averaged square of the regular wave
function on a circle with radius $a$, i.e.\ the boundary to the
fully chaotic phase space, yielding $\gamma_{mn}^{0} = N_{mn}^{2}
J_{m}(j_{mn}a)^2/2$. 
Surprisingly, it is just about a factor of
$2$ larger for the parameters we studied. 
In Ref.~\cite{BarBet2007}
a related quantity is proposed, given by the integral of the squared
regular wave function over the quarter circle with radius $a$. This
quantity, however, is too small by a factor of order $100$ for the
parameters under consideration.

The eigenvalues and eigenfunctions of the mushroom billiard are determined by
numerically solving the Schr\"odinger equation.
Because of its superior computational efficiency we have chosen to use the
improved method of particular solutions
\cite{BetTre2005,BarBet2007} allowing a determination of the energies $E$
with a relative error $\approx 10^{-14}$.
Analyzing avoided
crossings of a given regular state with typically 30 chaotic states
we deduce from Eq.~\eqref{eq:ensembleaverage} the tunneling rate.
Note, that some pairs of regular states are very close
in energy, e.g.\ $E_{20,1}-E_{16,2}\approx 10^{-4}$, such that their avoided
crossings with a chaotic state overlap, making a numerical
determination of the smaller tunneling rate unfeasible within
the presented approach. Fig.~\ref{fig:rates_low_E} shows the tunneling rates $\gamma_{mn}$
for fixed radial quantum number $n=1,2,3$ and increasing azimuthal quantum
number $m$, comparing the theoretical prediction,
Eq.~\eqref{eq:final_formula}, with numerical results.
We find excellent
agreement for tunneling rates $\gamma_{mn}$ over $18$ orders of magnitude.

In the experiment the unavoidable coupling to the environment dominates for 
small tunneling rates, but Fig.~\ref{fig:rates_exp} shows, that in the microwave 
experiment the coupling by the antenna is negligible over three orders of 
magnitude. This is a promising aspect for future experimental studies of more 
complex systems, in particular when numerical and theoretical results
are not available.

In summary, we have presented an experimental, numerical,
and theoretical investigation of tunneling rates in the mushroom billiard.
We find agreement without any free parameter, which is unprecedented for billiards.
This success of the approach using a fictitious integrable system gives
confidence that in the future it can be applied to generic billiards, where the
determination of a suitable $\Hreg$ is more challenging.

We thank the DFG for support within the Forschergruppe 760
``Scattering Systems with Complex Dynamics'', the cooperation program between
the Universities of Marburg and Maribor,
the Ministry of Higher Education, Science and Technology of the
Republic of Slovenia, and R.~K. thanks the Kavli Institute for
Theoretical Physics at UCSB (NSF Grant No. PHY05-51164).

\end{document}